\documentclass[aip,apl,numerical,reprint,superscriptaddress,showpacs,
showkeys]{revtex4-1}
\usepackage{bm,graphicx,graphics,amsmath,amssymb,bm,epsfig,color}
\usepackage{euscript,tabularx} \usepackage{textcomp} \usepackage{longtable}

\begin{document}

\title{Temperature dependence of microwave voltage emission associated to spin-transfer induced vortex oscillation in
magnetic tunnel junction} 

\author{P. Bortolotti} \altaffiliation{Corresponding author. Electronic address: paolo.bortolotti@thalesgroup.com}

\affiliation{Unit\'e Mixte de Physique CNRS/Thales and Universit\'e Paris Sud 11, 1 av. Fresnel, 91767 Palaiseau, France}

\author{A. Dussaux} \affiliation{Unit\'e Mixte de Physique CNRS/Thales and Universit\'e Paris Sud 11, 1 av. Fresnel, 91767 Palaiseau, France}

\author{J. Grollier} \affiliation{Unit\'e Mixte de Physique CNRS/Thales and Universit\'e Paris Sud 11, 1 av. Fresnel, 91767 Palaiseau, France}

\author{V. Cros} \affiliation{Unit\'e Mixte de Physique CNRS/Thales and Universit\'e Paris Sud 11, 1 av. Fresnel, 91767 Palaiseau, France}

\author{A. Fukushima} \affiliation{National Institute of Advanced Industrial Science and Technology (AIST), Spintronics Research Center}

\author{H. Kubota} \affiliation{National Institute of Advanced Industrial Science and Technology (AIST), Spintronics Research Center}

\author{K. Yakushiji} \affiliation{National Institute of Advanced Industrial Science and Technology (AIST), Spintronics Research Center}

\author{S. Yuasa} \affiliation{National Institute of Advanced Industrial Science and Technology (AIST), Spintronics Research Center}

\author{K. Ando} \affiliation{National Institute of Advanced Industrial Science and Technology (AIST), Spintronics Research Center}

\author{A. Fert} \affiliation{Unit\'e Mixte de Physique CNRS/Thales and Universit\'e Paris Sud 11, 1 av. Fresnel, 91767 Palaiseau, France}


\begin{abstract} The temperature dependence of a vortex-based nano-oscillator induced by spin transfer torque (STVO) in magnetic tunnel junctions (MTJ) is considered. We obtain emitted signals with large output power and good signal coherence. Due to the reduced non-linearities compared to the uniform magnetization case, we first observe a linear decrease of linewidth with decreasing temperature. However, this expected behavior no longer applies at lower temperature and a bottom limit of the linewidth is measured.

\end{abstract}

\maketitle

A spin polarized current can exert a large torque on the magnetization of a
ferromagnet through a transfer of spin angular momentum~\cite{Stiles-2006}. This
mechanism offers a new method to manipulate a magnetization, and potentially a
stable precession can be reached~\cite{Kiselev-2003}. By converting such
dynamics into a high frequency voltage oscillation through the magnetoresistance
effect, the concept of spin transfer-torque nano-oscillator (STNO) has been
proposed as a promising device for new technological applications in
IC-Technologies. However, despite compelling
progresses\cite{Deac-2008,Houssameddine-2008,Georges-2009,Pribiag-2007}, present challenges
remain to both increase the output power and improve the coherence of the emitted
signal. Recently, we demonstrated \cite{Dussaux-2010,Khvalkovskiy-2009} that, by
considering a magnetic tunnel junction (MTJ) with a vortex ground state (free
layer), the so-called \emph{spin transfer-torque vortex oscillator} (STVO), we
obtain large integrated power ($P_{int} \simeq 5$ nW) for a small linewidth,
i.e., full width at half maximum, $\Delta f < 1$ MHz. However, although $\Delta
f$ is considerably reduced compared to the uniform magnetization case, a major
issue is still its origin, and consequently the origin of the phase noise. 

When the efficiency of the spin transfer-torque induced by $I_{dc}$ exceeds a
critical value, the vortex core starts to precess and eventually reaches a
stable gyrotropic motion. The frequency of such gyrotropic vortex mode~\cite{Guslienko-2008} is well-separated from the frequencies of others modes (radial and azimuthal spin waves).
This fact, in principle, should allow to avoid the excitation of multiple modes
and simplify the shape of the oscillation peak, i.e., reduces $\Delta
f$~\cite{Houssameddine-2009}. From non-linear oscillation models
\cite{Kim-2008,Slavin-2009}, $\Delta f$ is expected to be proportional to
the temperature T and to the nonlinearity of the system, i.e., 
\begin{equation} \label{eq: Linewidth} 
\Delta f = \left( 1+ {\nu}^2\right) \; \Gamma_G \; {{k_B T} \over {E(p)}} 
\end{equation}
Here, $\nu \propto N$, which is the non-linear frequency shift coefficient $N=df/dp$, $\Gamma_G$ is the intrinsic damping coefficient and $E(p)$
represents the energy of the auto-oscillation with power $p$. In the case of gyrotropic vortex oscillators, $N$ is weak compared to MTJs with uniform magnetization \cite{Houssameddine-2008}. Hence, we expect a quasi-linear dependence of $\Delta f$ with temperature. 

The samples are circular MTJs of 300 nm diameter, with a 10 nm NiFe thick free layer and a synthetic antiferromagnet (SAF) that acts as in-plane uniformly magnetized spin polarizer. The complete structure (with thickness in nm) is: PtMn(15) / CoFe(2.5) / Ru(0.85) / CoFeB(3) / MgO(1.075) / NiFe(10) / Ta(7) / Ru(6) / Cr(5) / Au(200). Here, we focus on a MTJ pillar with resistances at room temperature (RT) $R_{P}= 49$ $\Omega$ for the case of two parallel uniform magnetizations, and $R_{AP}= 58$ $\Omega$ for two antiparallel uniform magnetizations, corresponding to a tunnel magnetoresistance ratio (TMR) of $\simeq 18$\%. The TMR ratio decreases down to 11\% for the maximum positive applied current ($I_{dc}=7$ mA), a standard behavior of the TMR bias dependence~\cite{Tsymbal-2003,Mistral-2006}. Note that positive current corresponds to electrons flowing from the free to the SAF layer. As concerns the temperature dependence, the TMR ratio increases up to 27\% at $20$ K which, again, corresponds to standard increasing ratios~\cite{Tsymbal-2003}. Several samples from the same wafer were measured and similar resistance values were obtained. For zero (or low) in plane field, the remanent magnetic configuration is a vortex state. In such system with a vortex and a uniform SAF polarizer, we have shown that a large output signal can be obtained when the perpendicular component of the spin polarization $p_z$ is both parallel to the core polarity (for positive current sign) and sufficiently large~\cite{Dussaux-2010,Khvalkovskiy-2009}. These two conditions are achieved, with the MTJ studied here, by applying a perpendicular field $H_{perp}>3.5$ kG. 

\begin{figure}[] \begin{center}
\includegraphics[width=0.45\textwidth]{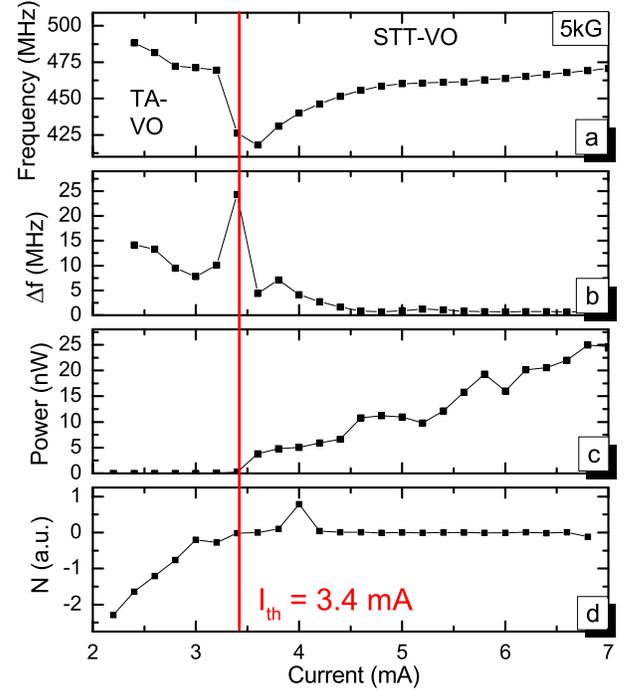} \end{center}
\caption{\label{fig:Critical_values} \small (a) Frequency $f$, (b) Linewidth $\Delta f$, (c) Integrated power $P_{int}$ and (d) non-linear coefficient $N=df/dp$ \emph{vs} $I_{dc}$ with $H_{perp}=5$ kG at $20$ K. Red vertical lines represents the threshold current value separating thermal activated vortex oscillations `\emph{TA-VO}' and vortex oscillations induced by spin-transfer torque `\emph{STT-VO}'.} 
\end{figure}

In Fig.~\ref{fig:Critical_values}, we report on the main features of the microwave signal associated to the spin-transfer induced gyrotropic motion of the vortex core. From the evolution with the current $I_{dc}$ of the four parameters, that are (a) the frequency $f$, (b) the linewidth $\Delta f$, (c) the integrated power $P_{int}$ and (d) the non-linear coefficient $N$, two regimes in the vortex dynamics can be clearly identified. Below a threshold current $I_{th}=3.4$ mA, marked by a red line, the microwave characteristics are associated to current induced thermally activated vortex oscillations (`\emph{TA-VO}') and $P_{int}$ is below $1$ nW. In this region, the trajectory of the vortex strongly depends on the disorder landscape, i.e., material defects and grains \cite{Min-2011}, which implies a complicated behavior for the frequency evolution with $I_{dc}$. Consequently, $\Delta f$ is large ($\simeq 10$ MHz, see Fig.~\ref{fig:Critical_values}(b)) and $N$, that is extracted from the derivative of $f$ and $P_{int}$ \emph{vs} $I_{dc}$, i.e., $N=df/dp=(df/dI)/(dp/dI)$ \cite{Slavin-2009}, varies strongly (see Fig.~\ref{fig:Critical_values}(d)). At the threshold current, $I_{th}$, $f$ undergoes a sharp change, $P_{int}$ increases rapidly to $3$ nW and $\Delta f$ has a maximum of $25$ MHz. This indicates the onset of spin-transfer torque induced vortex oscillations (`\emph{STT-VO}'). Then, we observe a transient region~\cite{Houssameddine-2009} for $3.5$ mA $<I_{dc}<4.5$ mA, in which $f$ increases rapidly and $\Delta f$ starts to decrease. Finally, above $I_{dc}>4.5$ mA, the frequency $f$ (Fig.~\ref{fig:Critical_values}(a)) follows a quasi-linearly behavior and the corresponding linewidth $\Delta f$ (Fig.~\ref{fig:Critical_values}(b)), reaches a minimum value ($\simeq 1$ MHz) that remains unchanged with increasing $I_{dc}$. The integrated power $P_{int}$ increases up to a value of $25$ nW for the maximum $I_{dc}$ (Fig.~\ref{fig:Critical_values}(c)). This behavior corresponds to an increase of the radius of the vortex-core oscillation. Interestingly, the non-linear coefficient $N$ is negligible for the whole current range. Note that, the maximum integrated power of all measured samples is 48 nW, which is, to our knowledge, the largest power value obtained with spin transfer nano-oscillators with $\Delta f$ of the order of $1$ MHz.

\begin{figure}[] \begin{center}
\includegraphics[width=0.40\textwidth]{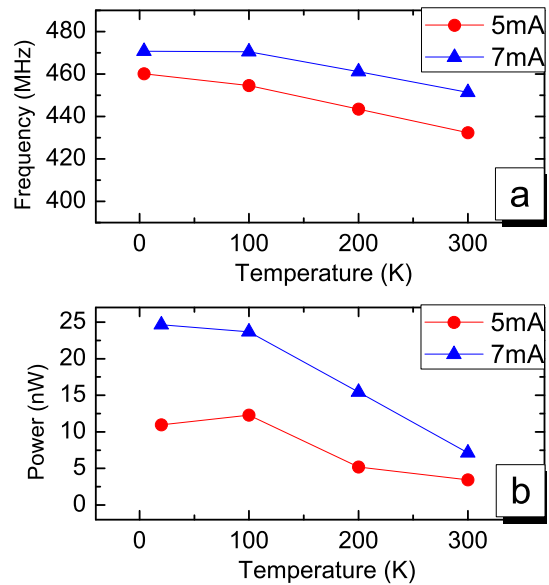} 
\end{center}
\caption{\label{fig:Temperature_A} \small The dependence of frequency $f$ (a) and integrated power $P_{int}$ (b) on temperature for $I_{dc}=$ $5$ and $7$ mA.} 
\end{figure}

Hereafter, we investigate the temperature dependence of the microwave characteristics in the (`\emph{STT-VO}') regime. In Fig.~\ref{fig:Temperature_A}(a), we focus on $f$ and $P_{int}$ for the case of $H_{perp}=5.5$ kG at four different temperatures (20-100-200-300 K). We see that for both $I_{dc}=5$ and $7$ mA, $f$ decreases by less than 5\% between $20$ K and $300$ K. We have checked that this reduction is related to the decrease in temperature of the saturation magnetization $M_s$. In order to get this latter dependance~\cite{Loubens-2009}, we have fitted the evolution of $f$ \emph{vs} $H_{perp}$ at each temperature (not shown) using the analytical expression valid for the gyrotropic motion of the vortex core: $f(H_{perp})=f(0)(1+H_{perp}/4 \pi M_s)$~\cite{Loubens-2009}. In Fig.~\ref{fig:Temperature_A}(b), we plot $P_{int}$ \emph{vs} T for the same current values. For both $I_{dc}=5$ mA and $I_{dc}=7$ mA, the integrated power $P_{int}$ decreases about $70$ \% between $20$ K and $300$ K. $P_{int}$ is expressed by the formula: 
\begin{equation} \label{eq: Power} 
P_{int}={\left({Z_0} \over{Z_0+R}\right)^2} \; \left( \Delta R_{osc} \; I \right)^2 
\end{equation}
where $R$ is the sample resistance and $Z_0$ is the circuit load (here $50$ $\Omega$). For the case of vortex core gyration, the amplitude of resistance oscillations is written as $\Delta R_{osc}=\left(\Delta M/M_s\right) \left(\Delta R_{TMR}/2\right)$ with $\Delta R_{TMR}$, the resistance variation between parallel and antiparallel configuration. The amplitude of magnetization oscillations is proportional to the radius of the vortex core $\Delta M/M_s \propto \; \rho/D$ where $D$ is the pillar diameter ~\cite{Guslienko-2002}. The temperature dependence of several parameters, i.e., $R$, saturation magnetizations of both SAF and free layer, $\Delta R_{TMR}$ and both frequency and radius of the oscillation, make difficult the detail explanation of the reduction of $P_{int}$ with temperature. However, a resonable estimation (60 \%) can be obtained by simply considering the temperature dependence of both $\Delta R_{TMR}$ and $M_s^{free}$. 

\begin{figure}[] \begin{center}
\includegraphics[width=0.40\textwidth]{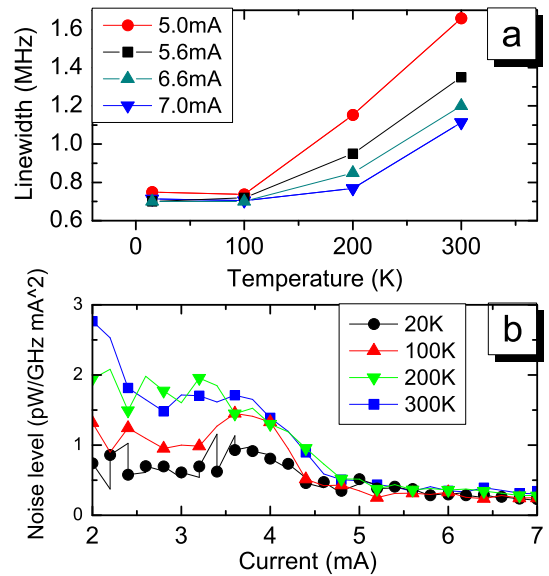} 
\end{center}
\caption{\label{fig:Temperature_B} \small (a) The dependence of the linewidth $\Delta f$ on temperature for $I_{dc}=$ ($5$, $5.6$, $6$ and $7$ mA). (b) Background noise $P_n$ \emph{vs} $I_{dc}$ for $20$ K, $100$ K, $200$ K and $300$ K.} 
\end{figure}

In Fig.~\ref{fig:Temperature_B}(a), we display the temperature dependence of $\Delta f$ at several current values, all above the threshold current. At each temperature and for each $I_{dc}$, the plotted value of $\Delta f$ corresponds to the average value of the peak linewidth measured in the $H_{perp}$ region for which the integrated power $P_{int}$ is maximum (see Fig. 2 in \cite{Dussaux-2010}). The most striking result is that we measure a constant bottom limit value $\Delta f \simeq 700$ kHz at low temperature, that is moreover almost independent of $I_{dc}$. Notably, it excludes the Joule heating as possible origin of this linewitdh limit. This is in contradiction to the expectation that in our weakly non linear vortex based oscillators, the linewidth should depend linearly on temperature. Indeed we recover this linear dependence of $\Delta f$ \emph{vs} $T$ above $100$ K for $I_{dc} = 5$ mA and at higher temperature for higher $I_{dc}$. A similar behavior, already observed in metallic nanocontact devices\cite{Schneider-2009}, was never measured in TMR structures, due to their intrinsic larger noise\cite{Georges-2009}. At $T=300$ K, the minimum $\Delta f = 1$ MHz measured for $I_{dc} = 7$ mA is indeed about twice larger than the one that can be estimated from Eq.~\ref{eq: Linewidth}. This difference, as well as the limit value of $\Delta f$ found in the temperature dependence, indicates that a new source of linewidth has to be considered in case of our vortex based oscillators. In order to get insights, we compare the background noise $P_n$ at different $I_{dc}$ and T (Fig.~\ref{fig:Temperature_B}(b)). $P_n$ was extracted for each spectra in the range between $100$ and $1200$ MHz. In the subcritical current regime, $P_n$ strongly depends on temperature and is constant while $I_{dc}$ increases. This behavior is expected in this regime of thermally actived vortex oscillations. On the contrary, above the threshold current, we first observe an intermediate regime in which $P_n$ decreases very fast to a bottom limit value, reached at the same $I_{dc}$ for all temperatures. Then, for $I_{dc}> 4.6$ mA, the background noise is constant. Indeed, this behavior is very similiar to the one observed for $\Delta f$ as function of $I_{dc}$, shown in Fig.~\ref{fig:Critical_values}(b). Interestingly, note that $P_n$, in this over-critical regime, is independent on temperature. These results clarify that, unlike the case of spin-transfer excitation of uniform magnetization in MTJ~\cite{Georges-2009}, the spin transfer torque is not the main source of the linewidth through the non linearity of the frequency in our STVO. Furthermore, the comparison of our quality factor $Q = f/\Delta f$, i.e., $250$ at RT and $650$ at $20$ K, in the overcritical regime to other $Q$ factors for vortex oscillators in very different sample geometries and materials~\cite{Mistral-2008,Pribiag-2007,Schneider-2009}, indicates that the observed limit value is not directly related to the entire magnetic volume but rather due to the dynamics in the region of the vortex core. Indeed, the classical formula (see Eq. \ref{eq: Linewidth}) for a (weakly) non linear oscillator~\cite{Slavin-2009,Kim-2008}, should be modified by adding an extra constant term $c_1$: $\Delta f = c_0 \; k_b T + c_1$. 

Here, we have considered only the action of the spin transfer torque on the lowest energy mode of a vortex, i.e., the gyrotropic motion of the vortex core in the NiFe layer. However in the actual structure, the magnetization of this layer is slightly coupled to the magnetization of the top layer of the SAF layer that might generate some additional noise due to spin transfer torque inside the SAF. In addition, even if we were not able to record any microwave peak at higher frequency related to other higher order spin waves modes of the vortex, we cannot exclude an eventual coupling between the vortex core gyration and these extra modes. This effect might be even stronger given that a large $H_{perp}$ is applied. An other possible mechanism of noise would be the existence of local inhomogeneties of the effective field that would be seen by the vortex core during its motion as a fluctuating field. These possible explanations need further investigations through micromagnetic simulations and time-domain measurements~\cite{Pribiag-2009,Keller-2010}. 

In conclusion, we find that, in the regime of large vortex trajectories, the non linear coefficient is much lower than in uniform  magnetization. It explains why the peak linewidth is almost constant on the whole current range. We investigate the origin of the linewidth by measuring the temperature dependence of the rf signal. We find that by decreasing the temperature, the linewidth first linearly decreases as expected but then saturates and reaches a bottom limit value (about $700$ kHz) independent on $I_{dc}$. Thus, we conclude that both thermal effects and spin transfer non linearities can not account for the observed results and an additional source of phase noise has to be invoked.

The authors acknowledge N. Locatelli for fruitful discussion, Y. Nagamine, H. Maehara and K. Tsunekawa of CANON ANELVA
for preparing the MTJ films and the financial support from ANR agency (VOICE PNANO-09-P231-36) and EU grant (MASTER No. NMP-FP7-212257).

\end{document}